\documentclass[12pt]{article}
\usepackage{amsmath}
\usepackage{amssymb}
\usepackage{graphicx}
\usepackage{subfig}
\usepackage{wrapfig}
\usepackage{fullpage}

\def\PROG#1{$\mathcal{#1}$}
\def\Proof{\par\noindent{\bf Proof:}\indent}
\def\QED{\hfill$\Box$\par\vskip1em}

\newtheorem{lemma}{Lemma}
\newtheorem{proposition}{Proposition}
\newtheorem{theorem}{Theorem}

\begin{document}
\title{Concurrent Geometric Multicasting}
\author{Jordan Adamek$^*$ \and Mikhail Nesterenko$^*$ \and James Scott Robinson$^*$ \and
        S\'{e}bastien Tixeuil$^\star$}
\date{$^*$Kent State University\\
  $^\star$UPMC Sorbonne Universit\'{e}s \& IUF}

\maketitle
\pagestyle{plain}
\begin{abstract}
We present \PROG{MCFR}, a multicasting concurrent face routing algorithm that uses geometric routing to deliver a message from source to multiple targets. We describe the algorithm's operation, prove it correct, estimate its performance bounds and evaluate its performance using simulation. Our estimate shows that \PROG{MCFR} is the first geometric multicast routing algorithm whose message delivery latency is independent of network size and only proportional to the distance between the source and the targets. Our simulation indicates that \PROG{MCFR} has significantly better reliability than existing algorithms. 
\end{abstract}

\section{Introduction} 

\noindent\textbf{Geometric routing.}
Geometric routing is transmitting a message from the sender to the targets on the basis of the geometric locations of the nodes. Geometric routing may offer a number of attractive features: it does not require nodes to maintain routing information beyond its immediate neighborhood; it can be stateless when no information is retained at the node as it forwards the message; its message size can be kept constant as each message carries limited amount of data, independent of network size.

\noindent \textbf{Unicasting face routing.} In \emph{unicasting}, a single source sends a message to a single target whose coordinates are known to the source. The simplest form of unicast geometric routing is greedy. In \emph{greedy routing}~\cite{Finn}, each intermediate node forwards the message to
its neighbor that is the closest to the target. Pure greedy
routing fails if the message encounters \emph{local minimum}: a node that
does not have neighbors closer to the target. A \emph{sequential} geometric routing algorithm, such as the classic GFG/GPSR~\cite{GFG,GPSR}, routes a single message in the \emph{greedy mode} until a local minimum is
encountered. The algorithm then switches to \emph{recovery mode} which
involves traversing the faces of a planar subgraph of the original
communication graph. Specifically, the algorithm traverses the faces that intersect the line that connects the source and the target. 

There are several disadvantages of sequential face traversal of GFG and similar algorithms. A sequential algorithm
cannot predict which traversal direction results in shorter distance. Hence, in the worst case, the latency of message delivery of such algorithms is proportional to the network size. Furthermore, such algorithms have low reliability: a single transmission fault results in complete message delivery failure. Last, sequential algorithms are unable to determine if the target is disconnected from the source, possibly resulting in the message remaining in the network arbitrarily long.
A \emph{concurrent} unicast algorithm CFR~\cite{CFR} sends a pair of messages to traverse the source-target line concurrently. This naturally selects the shortest face traversal direction.  Moreover, the second message provides greater robustness in case of message loss. The meeting of these messages signifies the end of face traversal and automatically determines whether the target is connected to the source. These advantages are offset by greater message cost.

\noindent\textbf{Multicasting.} Geometric multicasting requires the source to transmit the same message to several targets. The source knows the coordinates of the targets. 
Unicasting to each target separately may be inefficient if several targets are located near each other since multiple redundant messages are sent to similar locations. Efficient multicasting optimizes the routes the messages take so that a single
message is routed to multiple targets as long as possible.
LGS~\cite{LGS} computes a minimum Euclidean length spanning tree rooted in the source and containing all the targets. LGS then uses geometric unicast to transmit the messages along the edges of this tree. More sophisticated schemes of PBM~\cite{PBM} and GMP~\cite{GMP} compute Euclidean Steiner tree\footnote{A Euclidean Steiner
tree is a minimum length spanning tree with virtual
\emph{Steiner} nodes that connects the source and the destinations.} at every intermediate node and then unicast the message along this tree. 
This scheme improves efficiency since the total length of Steiner tree edges may only be
half that of the minimum spanning tree~\cite{steiner}.  In both GMP and PBM, the message is
geometric unicast towards the node, real or virtual, that roots the
subtree of the target destinations. As the message approaches a
virtual node, it may be advantageous to \emph{split} the message and route
separate messages towards different destination groups. GMP is shown to be more efficient than PBM in route selection and Steiner tree computation~\cite{GMP}.
Overall, all these multicasting algorithms use sequential face routing to recover from a
local minimum. Hence, they are prone to inefficient route selection, message loss and inability to detect network disconnection.

%\ \\
\noindent\textbf{Our contribution.} In this paper, we present \PROG{MCFR}: the first concurrent multicasting face routing algorithm. The algorithm computes Steiner tree of the targets and then concurrently routes the message around all the faces that intersect this Steiner tree. We prove it correct, provide asymptotic measures of its latency and message complexity and then evaluate its performance through simulation. We show that, unlike sequential multicasting algorithms, the latency of \PROG{MCFR} is independent of network size. It only depends on the distance between the source and the target. Our simulation shows that due to concurrency, the delivery ratio of \PROG{MCFR} significantly exceeds that of sequential algorithms.

\section{Notation and Definitions} 

\noindent\textbf{Wireless network, message and node limitations.} A \emph{wireless network} $G = (N, E)$ is represented as a graph where $N$ is a set of nodes that are devices capable of exchanging messages wirelessly, while $E$ is a set of edges connecting the nodes if the two adjacent nodes can send messages directly. Two such nodes are \emph{neighbors}. The communication is bi-directional and the graph is undirected. Every node has unique planar coordinates that \emph{embeds} the graph into the geometric plane. When it is clear from the context, we refer to a graph's embedding as just graph. 

\noindent\textbf{Planarity, face traversal, Steiner tree.} A graph embedding is \emph{planar} if the graph edges intersect only at vertices. For short, we call this planar embedding of a graph, a \emph{planar graph}. A  \emph{connected planar subgraph} is a subset of vertices and their induced edges such that the resultant graph is planar and connected. Finding a maximum planar subgraph of a general graph is NP-hard~\cite{NPHARD}. However, for certain graphs, the task may be solved efficiently. A graph is \emph{unit-disk} if a pair of its vertices $u$ and $v$ are neighbors if and only if the Euclidean distance between them is no more than $1$. Such a graph approximates a wireless network. In such a graph, a connected planar subgraph may be found by local computation at every node using Relative Neighborhood or Gabriel Graph~\cite{GFG,GS69,GPSR,T80}.
\emph{Face} is a region of the plane such any two points of the region may be connected by a continuous curve that that does not intersect the edges of the graph. A planar embedding of a finite graph divides the plane into a finite set of faces. The areas of all but one of the faces are finite. The finite area faces are \emph{internal}. The infinite face is \emph{external}. For example, in Figure~\ref{figMulticastingExample}, the graph has three internal faces: $F$, $G$ and $H$ and the external face.

Consider node $u$ and its neighbors $v$ and $w$. Node $w$ is \emph{next-right} after $v$, if it is the next neighbor of $u$ after $v$ clockwise; it is \emph{next-left} after $v$, if it is next to it counter-clockwise.  For example, in Figure~\ref{figMulticastingExample}, $f$ is next-right neighbor of $i$ after $s$. Observe that if $w$ is next-left after $v$, then $v$ next-right after $w$. Node $u$, its two neighbors $v$ and $w$ and the two incident edges form \emph{angle} $\angle vuw$. An angle \emph{intersects} a segment of a line if at least one of its edges lies on or intersects this segment. For example, $\angle \textit{sif}$ intersects segment $sx$. Note that we limit angle intersection to the fixed-size graph edges, not the infinite half-rays of a classic geometric angle.
In a planar graph, to traverse a face, messages are routed using right- or left-hand-rule. In the \emph{right-hand-rule}, if a node receives a message, it forwards the message to the neighbor that is next-right after the sender. In the \emph{left-hand-rule}, the message is forwarded to the next-left neighbor. For example, in Figure~\ref{figMulticastingExample}, if $i$ receives a right-hand-rule traversal message from $s$, then it forwards it to $f$.  
Two messages are \emph{mates} if they are traversing the same face in the opposite directions. A single node is able to detect mates if the sender of each message is the receiver of the other and the traversal direction of the two messages is opposite.
A \emph{Euclidean minimum Steiner tree} connects a selected set of nodes on a plane, possibly with added \emph{virtual nodes}, by a graph of minimum total length. The problem of computing such tree is NP-hard~\cite{GJ79}. GeoSteiner is the most successful algorithm that computes the exact solution in reasonable time~\cite{geosteiner}. Efficient polynomial approximations are also available~\cite{steinerApprox,improvedSteinerApprox}. For the rest of the paper, we refer to the Euclidean minimum Steiner tree as just Steiner tree and ignore the fact that it is being approximated.

\noindent\textbf{Message and node memory constraints.} To help with routing, a message carries routing information. Only \emph{constant size messages} are allowed. This means that the message may carry only a fixed number of node coordinates and related information. This fixed number is independent of the network size. This limitation, for example, precludes a routing algorithm from requesting the message to carry its entire route. Each message always carries the immediate sender, the node that transmitting this message, and immediate receiver, the node the message is being sent to. Each node stores the coordinates of its neighbors. No other information, either temporarily or permanently, can be stored by the node. This limitation precludes nodes from maintaining extensive routing tables of the network or storing state information between message transmissions.

%\ \\
\noindent\textbf{Steps, computations, fairness, multicasting}. Every node has a \emph{send queue} $SQ$ that collects messages to be sent. A message is transmitted by taking it from the sender's send queue, transferring it to the receiver and processing it according to the routing algorithm. In the theoretical discussion about the algorithm, we assume that this transferal and processing is done in a single atomic \emph{step}. The \emph{atomicity} of the step means that it may not overlap with the steps of this or any other nodes. In the simulation section, this assumption is lifted.
\emph{Computation} is a sequence of atomic steps that starts in an initial state of the algorithm. A computation is \emph{fair} if each message, in the send queue of every node, is either transmitted or removed from this queue. That is, a message may not ``get stuck'' in a send queue forever. We consider only fair computations. A computation with a finite number of steps is itself \emph{finite}. A routing algorithm is \emph{terminating} if its every computation is finite. A terminating routing algorithm never leaves messages circulating in the network indefinitely.
A \emph{multicasting routing algorithm} ensures a message is delivered from the \emph{source} to the set of \emph{targets}. The source knows the coordinates of the targets and these coordinates fit in single message.  For example, in Figure~\ref{figMulticastingExample}, source $s$ may need to send a messages to targets $b$, $d$ and $k$. To aid in navigation, a multicasting algorithm may compute a Steiner tree that includes source, targets and virtual nodes $x$ and $y$.

\begin{figure}[htb]
\scriptsize
\begin{tabbing}
123412345678901\=1234\=1234\=1234\=1234\=1234\=1234\=1234\=1234\=1234\=\kill
\>\textbf{node} $s$\\
\>\textbf{compute} $T$  \quad  /* $T$ is Steiner tree */ \\
\>\textbf{foreach} $\angle asb$  that intersects $T$ \textbf{do}\\
\>\>\textbf{add} $L(s,T,a)$ to $SQ$\\
\>\>\textbf{add} $R(s,T,b)$ to $SQ$\\
\\
\>\textbf{node} $n$\\
\>\textbf{if} \textbf{receive} $L(s,T,a)$ \textbf{then}\\
\>\>\textbf{if} $R(s,T,a) \in SQ$ \textbf{then} \\
\>\>\>/* found mate */\\
\>\>\>\textbf{discard} $R(s,T,a)$ \textbf{from} $SQ$\\
\>\>\textbf{else} \\
\>\>\> \textbf{if} $n \in T$ \textbf{then} \\
\>\>\>\>			\textbf{deliver} $L(s,T,a)$ to $n$ \\
\>\>\>/* let $b$ be the next left after $a$ */\\
\>\>\>\textbf{add} $L(s,T,b)$ to $SQ$ \\
\>\>\>\textbf{if} $\angle anb$ intersects $T$ \textbf{then} \\
\>\>\>\>/* split message */ \\
\>\>\>\>\textbf{foreach} $\angle cnd \neq \angle anb$ that intersects $T$ \textbf{do}\\
\>\>\>\>\>/* let $d$ be the next left after $c$ */ \\
\>\>\>\>\>\textbf{add} $R(s,T,c)$ to $SQ$\\
\>\>\>\>\>\textbf{add} $L(s,T,d)$ to $SQ$ \\
\>\textbf{if} \textbf{receive} $R(s,T,a)$ \textbf{then} \\
\>\>/* handle similar to $L(s,T,a)$ */
\end{tabbing}
\caption{\PROG{MCFR} pseudocode.} \label{figMCFRcode}
\end{figure}

\section{Algorithm Description}

The pseudocode of algorithm \PROG{MCFR} is shown in Figure~\ref{figMCFRcode}. It operates as follows. The sender computes the Steiner tree $T$ of targets and itself. For each angle that intersects $T$, it sends a pair of mates with right $R$ and left $L$ traversal directions. 
Every message carries the encoding of $T$. For simplicity, we assume that source itself is never a target.

Once some node $n$ receives a message, it checks if there is its mate in its send queue $SQ$. If the mate is found, both messages are removed and further processing stops as this completes the traversal of a face. If there is no mate, $n$ checks if it is the target and delivers the message. After the delivery, message processing continues. Specifically, the message is forwarded along the face that it traverses.  A node is \emph{juncture} if at least one of its angles intersects $T$. If $n$ is a juncture node then $n$ \emph{splits} the message by injecting a pair of mates in every angle that intersects $T$. Observe that the source node is always a juncture.

\begin{figure}[htbp]
  \centering
\begin{minipage}[t]{0.5\linewidth}
\includegraphics[width=\linewidth]{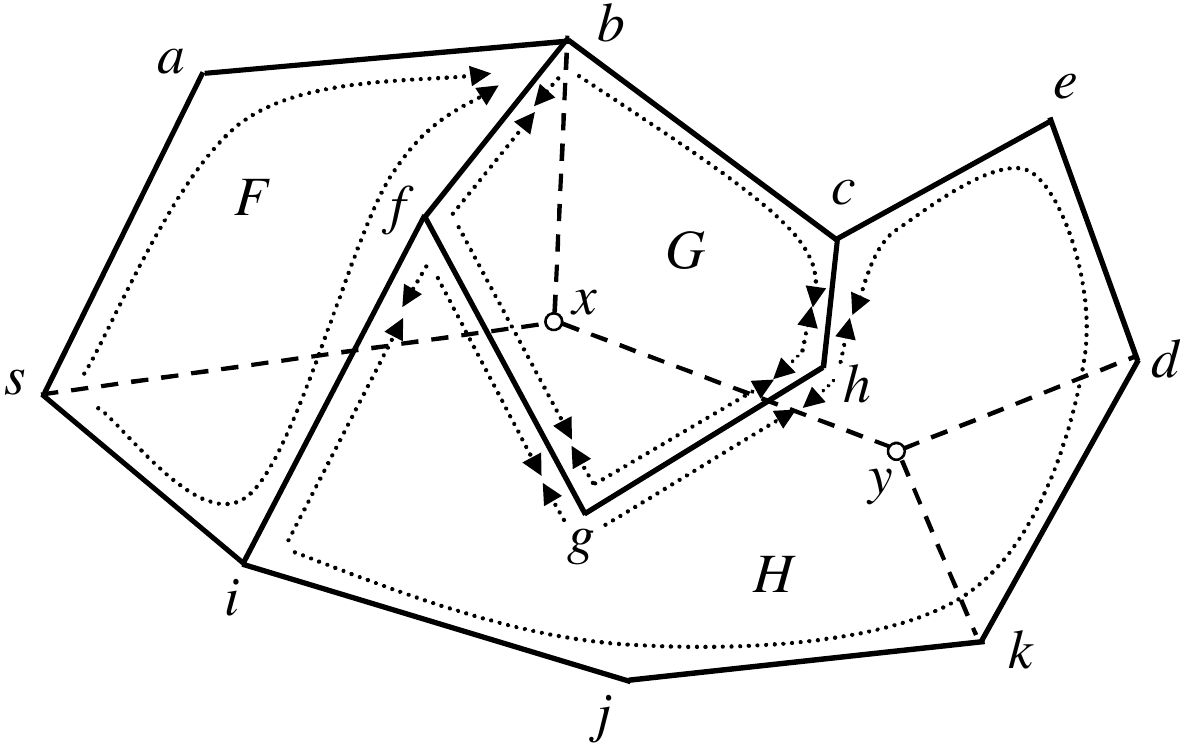}
\caption{\PROG{MCFR} operation example.}\label{figMulticastingExample}
\includegraphics[width=\linewidth]{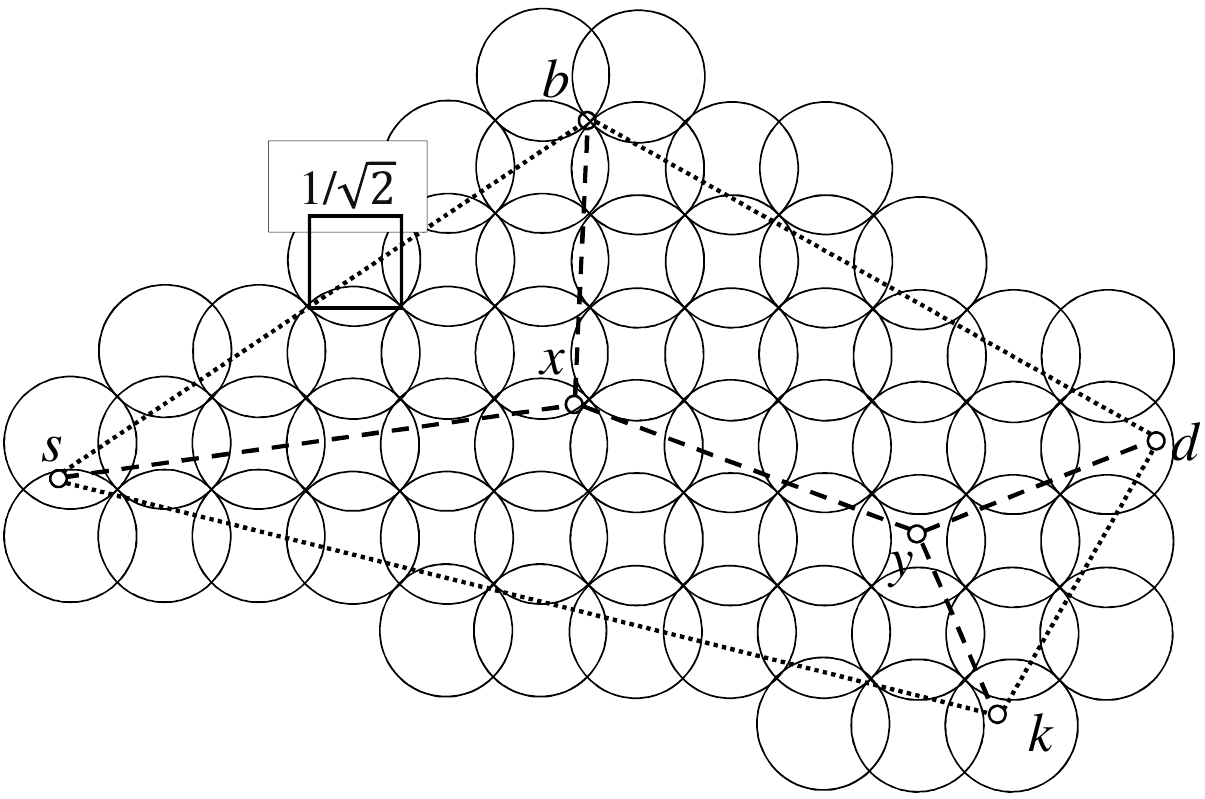}
\caption{Covering the source-target Steiner hull with unit circles. Illustration for the proof of Theorem~\ref{thrmMessageBound}.}
\label{figUnitDiskCover}
\includegraphics[width=\linewidth]{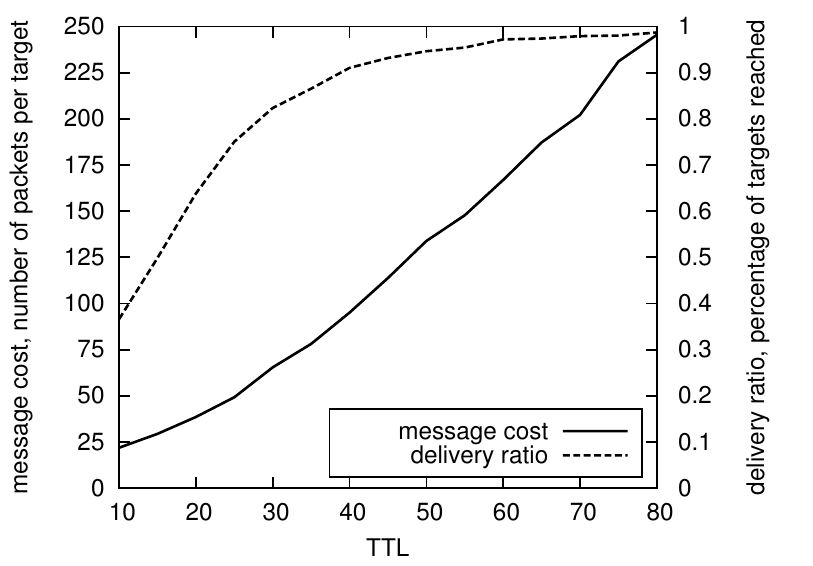}
\caption{Time to live evaluation for MCFR, Steiner Tree, 15 dBm.}\label{figTTL}
\end{minipage}
\end{figure}

Let us consider an example of \PROG{MCFR} operation shown in Figure~\ref{figMulticastingExample}. Source node $s$ computes the Steiner tree $T$, determines that $\angle isa$ intersects $T$ and sends a left-hand-rule $L$ traversing message to $a$, while sending $R$ to $i$. This injects a pair of mates into face $F$. When $a$ receives a message, it forwards it to its next-left left after $s$ neighbor $b$. When $i$ receives a message, it determines that $i$ is a juncture node and the angle that intersects $T$ is
$\angle \mathit{fij}$. Node $i$ injects a pair of mates by sending an $L$ to $f$ and an $R$ to $j$. Meanwhile, $i$ also forwards the originally received message to $f$. Node $f$ injects mates into faces $H$ and $G$ and forwards the original message to $b$. Node $b$ injects messages into face $G$ and forwards the message to $a$. However, then $b$ receives a message from $a$. Once $b$ receives a message from $a$, it inspects its $SQ$, finds its mate and destroys both messages. This completes the traversal of face $F$. The operation of \PROG{MCFR} continues until all faces that intersect $T$ are traversed.

\begin{figure}[htbp]
  \centering
%\hspace{0.02\linewidth}
\begin{minipage}[t]{0.5\linewidth}
\subfloat[15 dBm]{\includegraphics[width=\linewidth]{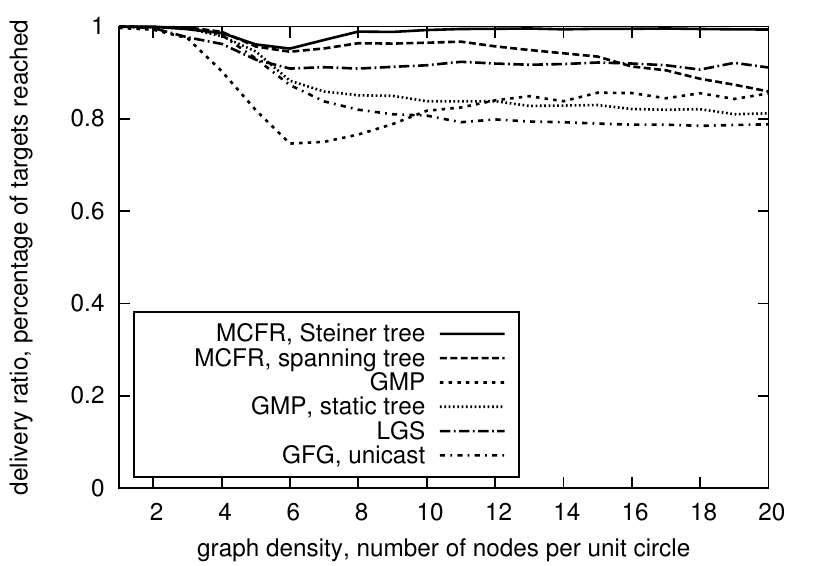}} \\
\subfloat[7 dBm]{\includegraphics[width=\linewidth]{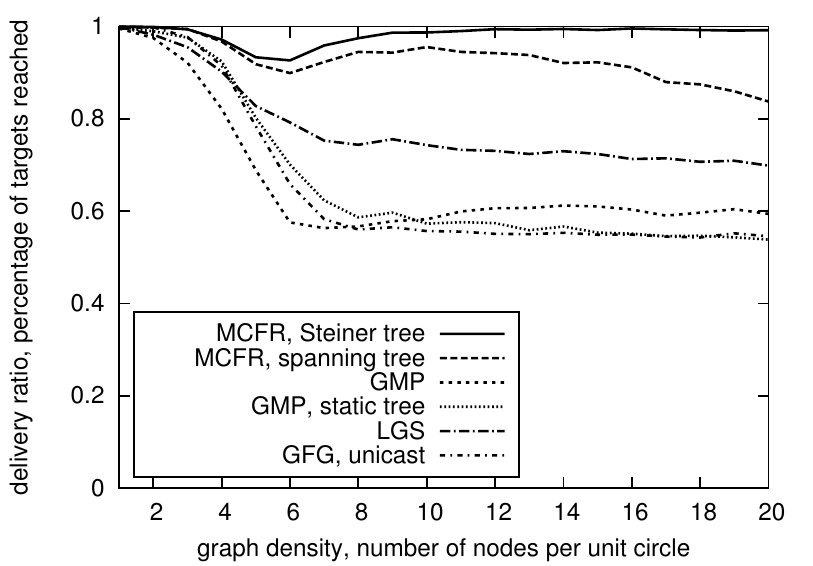}} \\
\subfloat[0 dBm]{\includegraphics[width=\linewidth]{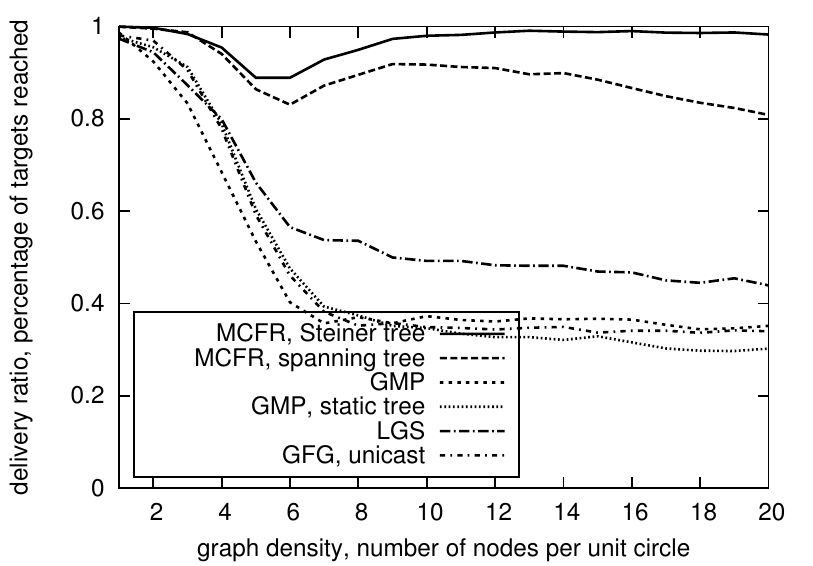}}
\caption{Delivery ratio.}\label{figDeliveryRatio}
\end{minipage}
\end{figure}

\section{Correctness Proof and Efficiency Bounds}

\textbf{Correctness proof.}
Let us introduce some notation to aid in the correctness discussion.
A node is \emph{segment-visited}, or just \emph{visited}, with respect to a particular face if it was visited during the traversal of this face. It is \emph{unvisited} otherwise. A \emph{visited segment} of a face is a sequence of neighbor nodes that are visited. A \emph{segment-border} of a visited segment, with respect to a particular face, is a visited node with a neighbor that is either (i) unvisited or (ii) carrying a message for the border node. A non-border visited node is \emph{segment-internal}. Note that an edge in a planar graph is adjacent to two faces. Thus, a node may be visited in one face but unvisited in another. Two faces are \emph{adjacent} if they  share a common juncture node. 
Two faces $F$ and $G$ are \emph{juncture connected} if there exists a sequence of adjacent faces that starts in $F$ and ends in $G$.  Observe that by the design of the algorithm, once the juncture is visited, it splits the message in every angle that intersects $T$. That is, a juncture is visited in every adjacent face at once.

\begin{lemma}\label{lemBorderMessage}
In \PROG{MCFR}, for every face $F$ with a visited segment, the segment border node has a message to send across to the unvisited neighbor. A segment-internal node never holds such a message.
\end{lemma}

\Proof
The proof is by induction on the nodes of a particular face $F$. A visited segment is created in $F$ when a juncture node is visited. This juncture may be the source or another node splitting the message when it is visited in an adjacent face. Once the visited segment is created, it contains a single border node with two messages sent in the opposite directions. This is our base case. 

Let us consider a computation of \PROG{MCFR} where every visited segment of every face is as stated in the conditions of the lemma.  First, let us consider a message transmission by node $u$ adjacent to face $F$. By the induction hypothesis, $u$ may only be a border node. The message recipient $v$ may be an unvisited node or a visited node that has a message for $u$. Let us consider the unvisited case first. Once unvisited node $v$ receives a message from $u$, $v$ becomes a new border node with this message while $v$ becomes segment internal without a message. Hence the conditions of the lemma hold. If $v$ is visited, then by definition of the border node, it holds a mate for the message that $v$ transmits. Moreover, by the condition of the lemma, $v$ is also a border node of an adjacent segment. Once $v$ receives the message from $u$, both mates are destroyed and both $u$ and $v$ become segment-internal nodes. This also preservers the condition of the lemma.

Let us now discuss the transmission of a message by a node $u$ that is not adjacent to $F$. The only way that it may affect $F$ is if $v$ is a juncture adjacent to $F$. However, by the design of the algorithm, the juncture is instantly visited in every adjacent face. That is, when $v$ receives a message transmission, it is not visited. Once it receives a message, it creates a new visited segment in $F$ with a single border node $v$ and appropriate outgoing messages. 

That is, regardless of the message transmissions, the conditions of the lemma are preserved.
\QED

\begin{lemma}\label{lemAllFaceVisited}
In \PROG{MCFR}, if a face has a visited segment, every node adjacent to this face is eventually visited and none holds messages. 
\end{lemma}

\Proof If a face with a visited segment contains an unvisited node, then, at least one such unvisited node is adjacent to a border of a visited segment. Due to Lemma~\ref{lemBorderMessage}, this border node has a message to be sent to the unvisited adjacent node. Since we only consider fair computations of the routing algorithm, this message is eventually transmitted. Once the message is sent, the adjacent node becomes visited. This process continues until all nodes of the face are visited. Once all nodes are visited they become internal and, according Lemma~\ref{lemBorderMessage}, do not hold message. Hence the lemma.\QED

\begin{lemma}\label{lemAllVisited}
In \PROG{MCFR}, if a face intersects the Steiner tree $T$, then every node adjacent to this face is eventually visited. 
\end{lemma}
\Proof Consider the face that contains the source node $s$. The algorithm starts by creating a single-node visited segment there. According to Lemma~\ref{lemAllFaceVisited}, every node in this face is eventually visited. This includes all junctures adjacent to this face. Repeated application of Lemma~\ref{lemAllFaceVisited} proves this lemma.
\QED

\begin{proposition}\label{propAllConnected}
If a target node is connected to the source node, then this node lies on a face that is juncture connected to the source node face. 
\end{proposition}

The below theorem follows from Proposition~\ref{propAllConnected} and Lemma~\ref{lemAllVisited}.

\begin{theorem}\label{thrmMCFRcorrect}
Algorithm \PROG{MCFR} guarantees termination and delivery of the message from the source to all targets connected to the source. 
\end{theorem}

%\ \\
\noindent\textbf{Efficiency bounds}.
\emph{Latency} of an algorithm is the shortest path that the message may take to reach the target. In multicasting, the latency is the longest such path among all targets. 
For latency estimation, following Kuhn et al~\cite{TheoryPractice}, we assume that the network graph $G$ is bounded degree since such graph can be efficiently obtained in a unit-disk graph by computing a connected-dominating set of $G$. 
The latency of a unicast concurrent face routing algorithm is established to be $O(t^2)$ where $t$ is the distance between the source and the target~\cite[Theorem 2]{CFR}. 

\begin{theorem}\label{thrmLatencyBound}
\PROG{MCFR} latency is in $O(d^2)$ where $d$ is the Steiner tree diameter. 
\end{theorem}

\Proof
Let $d$ be the diameter of the Steiner tree and $m$ is the number of multicast targets. A Euclidean Steiner tree has at most $m-2$ virtual nodes~\cite{steiner}.  In the worst case, the multicast message in \PROG{MCFR} has to sequentially reach all nodes in the Steiner tree. That is, the total latency is $O((2m - 2)d^2)$ which is $O(md^2)$. The theorem's claim follows if the number of targets is constant.
\QED

\emph{Message cost} of an algorithm is the total number of messages expended in delivering  it to the targets. In the worst case, \PROG{MCFR} traverses every face of the graph. An edge is adjacent to two faces, hence \PROG{MCFR} may send up to $2|E|$ messages where $E$ is the number of edges in the graph. However, for most graphs, \PROG{MCFR} is a lot more efficient. To give a more realistic message cost estimate of \PROG{MCFR} we make several assumptions about the network graphs. 

The graph is \emph{face smooth} if there
are two constants $c_1$ and $c_2$ that are independent of network
parameters such that (i) for each face $\rho ^2 < c_1 a$ where $\rho$
is the perimeter of the face, and $a$ its area, and (ii) for any two
points in the graph, $a_s < c_2 \frac{\pi d^2}{4}$ where $a_s$ is the
area of all internal faces that intersect the line between these two
points, and $d$ is the Euclidean distance between them. For an internal
face, the area computation is straightforward; for the external face,
an area of an arbitrary figure enclosing the graph, for example a
convex hull, is considered.  The first assumption places limits on how
``ragged'' the perimeter of the face may be, while the second limits
how ``uneven'' the faces may be in size by assuming that the area of
all intersecting faces is included in a certain disk whose diameter is
related to the distance between two devices.  These assumptions hold
for most realistic wireless communication graphs such as unit-disk
graphs. \emph{Steiner hull} is the convex hull that contains the nodes of the Steiner tree. It is known that virtual nodes are internal to the Steiner hull~\cite{steiner}.

\begin{theorem}\label{thrmMessageBound} For face smooth graphs, 
the message cost for \PROG{MCFR} is in $O(|H| + \sqrt{|G|})$, where $|H|$ is the area of the Steiner hull and $|A|$ is the area of the complete graph $G$.
\end{theorem}

\Proof
We completely cover the Steiner hull $H$ with unit-disks. See Figure~\ref{figUnitDiskCover} for illustration. In this arrangement, each unit disk covers a square with side length of $\frac{1}{\sqrt{2}}$. Since $k$ is the maximum node degree, the number of nodes in each unit disk is no more than $k$. Hence, the number of nodes inside $H$ is:  $k\frac{|H|}{(\frac{1}{\sqrt{2}})^2} = 2k|H|$.

Since there are at most $k$ neighbors, each node may be adjacent to at most $k$ edges. A message may be sent across each edge at most twice. Hence, the number messages to be sent inside $|H|$ is $4k^2|H|$ which is in $O(|H|)$.

Let us know estimate the number of messages it takes to traverse all the faces that intersect $H$. Any convex polygon can be inscribed into a rectangle whose area at most twice the size of that of the polygon~\cite{rectangleApproximation}. 
This means that $H$ can be inscribed into a rectangle whose area is at most $2|H|$. The perimeter of this rectangle is $4\sqrt{2|H|}$. We assume that the graph is face-smooth. Combining the two face smoothness conditions we obtain that the sum of the perimeters of all the internal faces that intersect $H$ has this relation:
	\[ p_s^2 = c_1 c_2 \frac{\pi (4\sqrt{2|H|})^2}{4} = c_1 c_2 \pi 4 \sqrt{2} |H|, \]
which means that $p_s$ is in $O(\sqrt{|H|})$.
Similarly, the perimeter of the external face is in $O(\sqrt{|G|})$. Since each face is traversed at most once, the total number of messages used to traverse internal faces that intersect $H$ as well as the external face is in $O(\sqrt{|H|} + \sqrt{|A|})$. Combined with the number of messages needed to traverse faces inside the Steiner hull we get $O(|H| + \sqrt{|H|} + \sqrt{|G|}) = O(|H| + \sqrt{|G|}).$ 
\QED

To summarize, Theorem~\ref{thrmLatencyBound} shows that the latency of message delivery of \PROG{MCFR} does not depend on the overall network size, just on the distance between the source and the targets; Theorem~\ref{thrmMessageBound} shows that the total number of messages sent by \PROG{MCFR} depends on the locations of the targets with respect to the source and the length of the external face of the graph. 

\begin{figure}
  \centering
\begin{minipage}[t]{0.5\linewidth}
\subfloat[15 dBm]{\includegraphics[width=\linewidth]{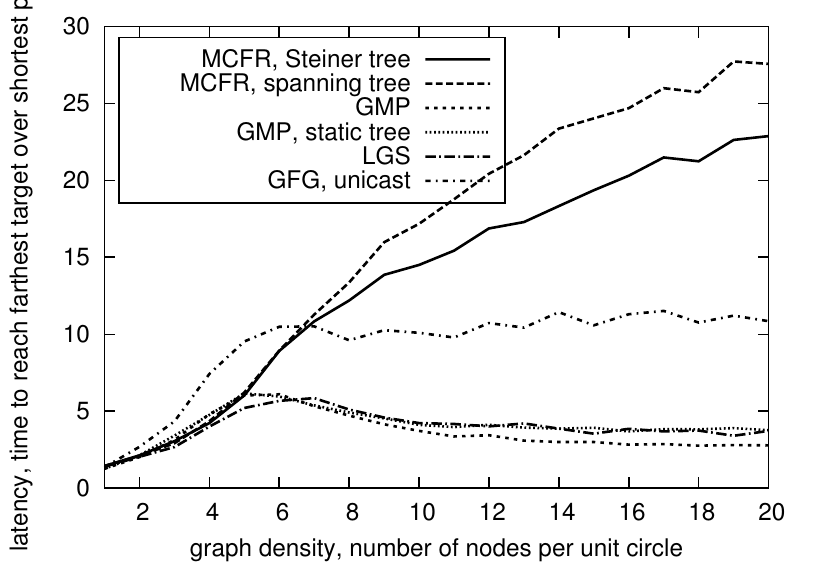}} \\
\subfloat[7 dBm]{\includegraphics[width=\linewidth]{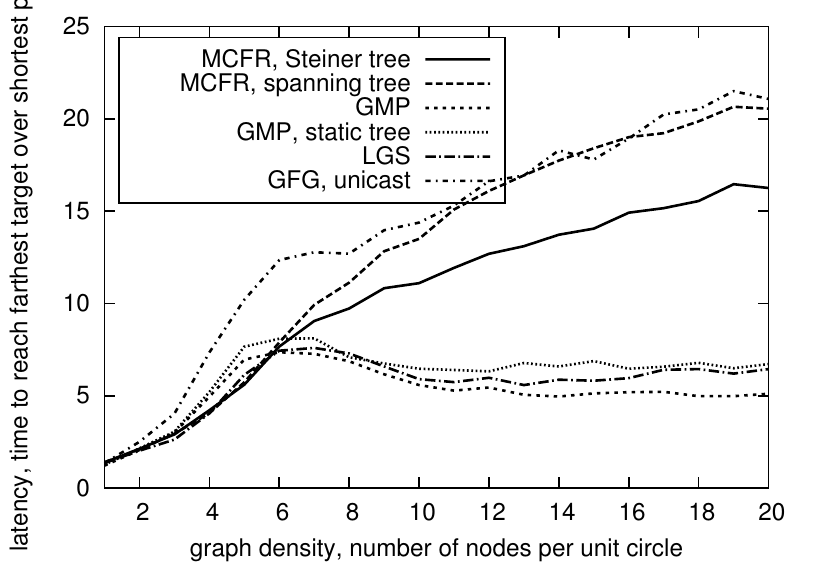}} \\
\subfloat[0 dBm]{\includegraphics[width=\linewidth]{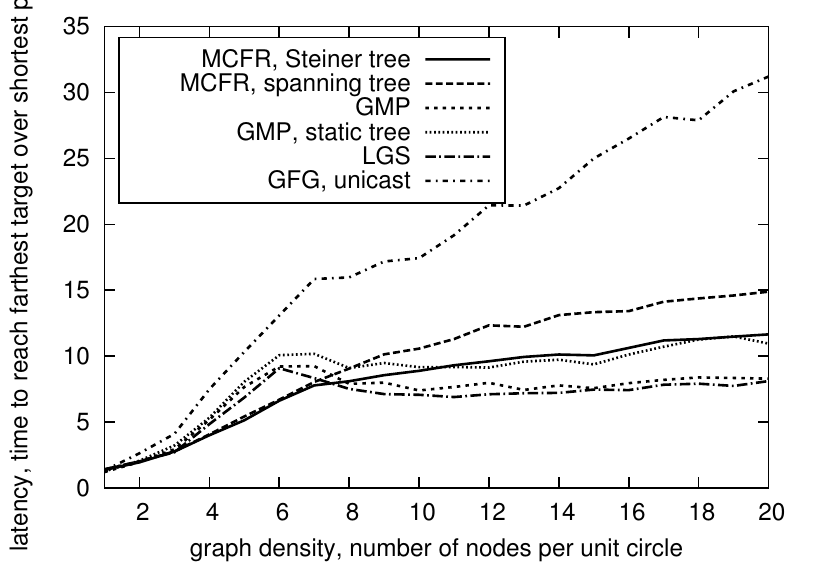}}
\caption{Latency.}\label{figLatency}
\end{minipage}
\end{figure}

\begin{figure}
  \centering
%\hspace{0.02\linewidth}
\begin{minipage}[t]{0.5\linewidth}
\subfloat[15 dBm]{\includegraphics[width=\linewidth]{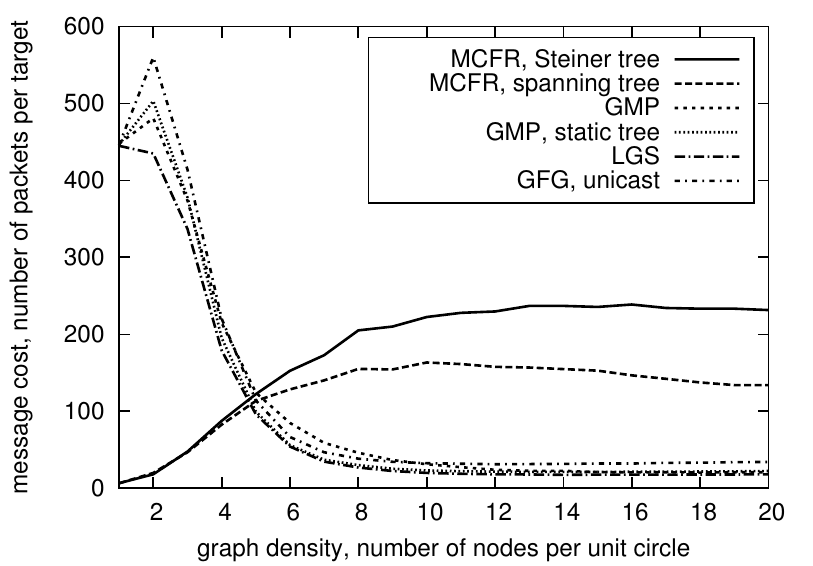}} \\
\subfloat[7 dBm]{\includegraphics[width=\linewidth]{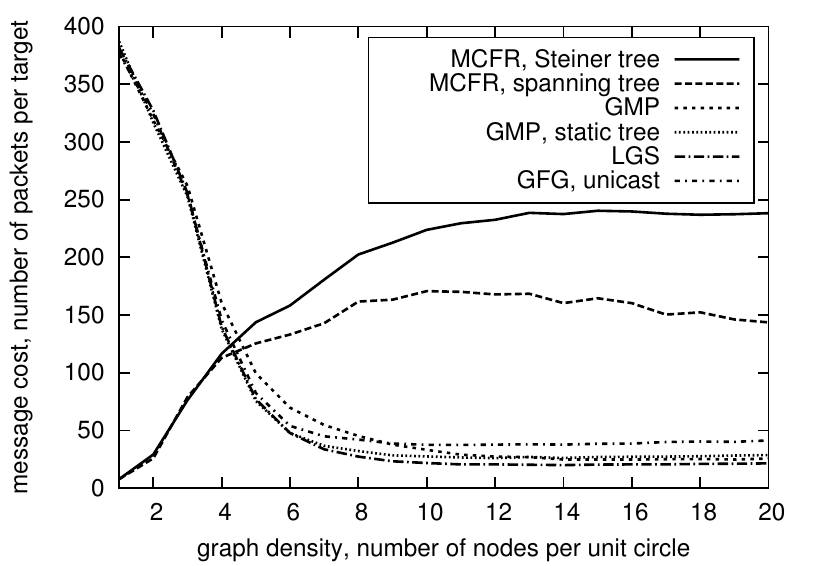}} \\
\subfloat[0 dBm]{\includegraphics[width=\linewidth]{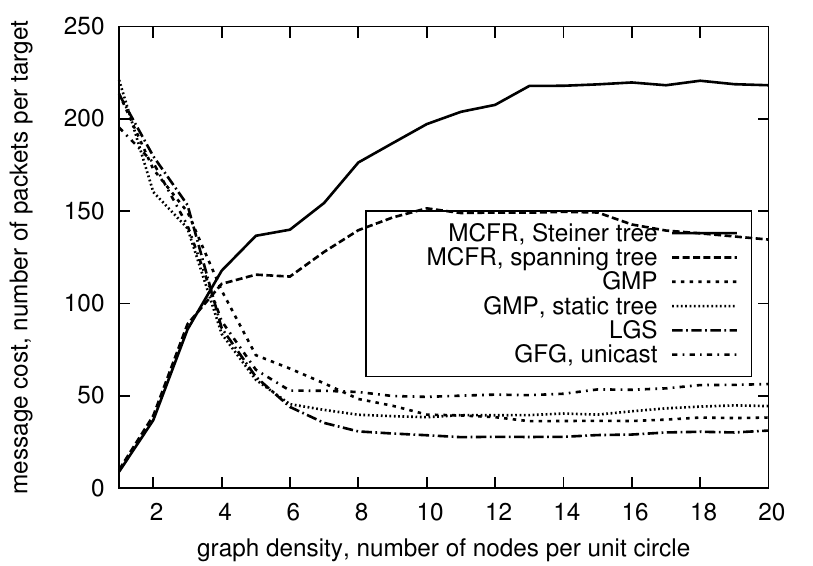}}
\caption{Message cost.}\label{figMessageCost}
\end{minipage}
\end{figure}

\section{Simulation}

\textbf{Setup.} To evaluate the performance of our algorithms, we implement them in WSNet~\cite{worldsensComplexity,DHM14C,DHMP14C,worldsensIPSN} wireless sensor network simulator. The simulated MAC layer is IEEE 802.15.4 with 866 MHz frequency band and BPSK modulation. The ratio model is freespace propagation with constant path loss and rayleigh fading~\cite{worldsensComplexity}. 
In our geometric simulation setup, we recreate and extend the setup of the classic geometric unicast algorithm study of Kuhn et al~\cite{TheoryPractice}. Specifically, we simulate $1000\times 1000$ meters field. The unit is $100$ meters. The field is populated by nodes placed uniformly at random to achieve a specific network density. For a fixed density, the number of nodes is calculated as follows. The total number of nodes $n$ is equal to the area of the field divided by the area of the unit circle and multiplied by the required density $d$. That is $n = d \frac{1000\times 1000}{\pi 100^2}$. To compute the network topology we use a unit-disk graph, then compute a Gabriel subgraph over it. The topology is calculated offline. We evaluate our algorithms' performance at three power levels: 15 dBm, 7 dBm and 0 dBm. The weaker the signal, the less reliable message transmissions are.  

\emph{Experiment} is a single delivery of a message from a particular source to a particular set of targets. In other words, it is a single complete computation of an algorithm. For each experiment, we generate a new random graph with randomly selected source and targets. The number of targets is selected to be $5\%$ of the total number of nodes. For each specific data point we conduct $1000$ experiments and compute the average value. 

We evaluate the performance of our algorithms according to three metrics. 
\emph{Delivery ratio} is the number of targets that receive the message divided by the total number of targets. Delivery ratio determines the reliability of the algorithm. \emph{Latency} is the the  time it takes the algorithm to deliver the message to the target that is geometrically furthest away from the source divided by the time it takes to unicast this message to this target using an optimum route. Latency determines the algorithm's speed of message delivery. \emph{Message cost} is the number of message transmissions divided by the number of targets. Since a message transmission is a radio broadcast, message broadcast to several neighbors is counted as a single transmission. The message cost counts all message transmissions regardless of delivery success. 
Raw latency and message cost do not take into account delivery success. For example, an algorithm that delivers only to the nearest target may have low latency and message cost.
To offset that, we divide the latency and message cost by the delivery ratio.

%\ \\
\noindent\textbf{Algorithms.} We simulate both sequential and concurrent multicasting algorithms. For sequential algorithms, as a baseline, we simulate the trivial algorithm that uses a unicast GFG to separately deliver the message to each individual target. We simulate LGS.  We simulate GMP as
presented in the original paper~\cite{GMP} as well as a variant of the algorithm where the Steiner tree is computed once at the source and is not recomputed by every intermediate node. 
For concurrent multicasting, we simulate two versions of MCFR. In the first version, MCFR navigates over Steiner tree; in the second, it routes over minimum Euclidean spanning tree. The trees are computed by the source.

In realistic environments, both concurrent and sequential algorithms have termination issues. 
If messages are lost, MCFR packets may not find mates. In this case, the messages may traverse graph faces indefinitely. On the other hand, a sequential algorithm does not detect if a target is disconnected. Again, a message for such disconnected target never reaches its destination. To force termination, we introduce time to live (TTL) for each message. A message is discarded after its TTL expires. To determine optimal TTL, we vary TTL then compute message cost and delivery ratio for MCFR with Steiner Tree and 15 dBm signal strength. The results are shown in Figure~\ref{figTTL}. The TTL of 55 hops seems to produce the best performance. For the rest of the experiments all algorithms have the TTL of 55. 

%\ \\
\noindent\textbf{Results and analysis.} The simulation results for delivery ratio, latency and message cost of the simulated algorithms are shown in Figures~\ref{figDeliveryRatio},~\ref{figLatency} and~\ref{figMessageCost} respectively. Let us discuss the results. The reliability of MCFR exceeds that of sequential multicasting algorithms. The gap widens as signal strength lowers and message loss increases. For $0$ dBm, the delivery ratio of MCFR Steiner Tree never drops below $90\%$ while the delivery ratio of most of the sequential algorithms goes below $40\%$. Interestingly, the reliability of simple unicasting to all targets exceeds that of GMP and LGS. This is due to the difference in their operation. To optimize message cost, GMP and LGS combine the messages to multiple targets as long as possible. However, a loss of such combined message results in failed delivery to all targets.

The latency of concurrent algorithms is higher than that of sequential algorithms. This is due to the greater number of messages contending for radio channel access. However, with lower signal strength and greater message loss, the gap narrows. For $0$ dBm, latency of concurrent and sequential algorithms is similar. The higher delivery ratio of concurrent algorithms match relative speed of delivery of the sequential algorithms. The only exception is the unicast GFG whose latency continues to be high. 

Concurrent and sequential algorithms exhibit different message cost dynamics. At lower densities greater number of targets is disconnected. This forces sequential algorithms to send aimless messages to wander around the network until their TTL expires. Concurrent algorithms do not have this issue. As the network density grows, the relative frugality of the sequential algorithms gives them message cost advantage over concurrent algorithms.

\section{Conclusion}
Algorithm~\PROG{MCFR} presented in this paper provides  good practical reliability and asymptotic latency that depends only on the Steiner tree diameter. As future work, we suggest the following. To further improve latency, apply this algorithm to minimum diameter Steiner trees~\cite{ding2014algorithms}. Evaluate fine-grained energy consumption of the algorihtms, for example using the approach of Bramas et al~\cite{BT16c}, as it may impact the survivability of strategic parts of the network.

%\bibliographystyle{plain}
%\bibliography{multicasting}

\begin{thebibliography}{10}

\bibitem{steinerApprox}
Sanjeev Arora.
\newblock Polynomial time approximation schemes for euclidean traveling
  salesman and other geometric problems.
\newblock {\em Journal of the ACM (JACM)}, 45(5):753--782, 1998.

\bibitem{worldsensComplexity}
Elyes Ben~Hamida, Guillaume Chelius, and Jean-Marie Gorce.
\newblock On the complexity of an accurate and precise performance evaluation
  of wireless networks using simulations.
\newblock In {\em Proceedings of the 11th international symposium on Modeling,
  analysis and simulation of wireless and mobile systems}, pages 395--402. ACM,
  2008.

\bibitem{GFG}
P.~Bose, P.~Morin, I.~Stojmenovic, and J.~Urrutia.
\newblock Routing with guaranteed delivery in ad hoc wireless networks.
\newblock {\em The Journal of Mobile Communication, Computation and
  Information}, 7(6):48--55, 2001.

\bibitem{BT16c}
Quentin Bramas and S{\'{e}}bastien Tixeuil.
\newblock Benchmarking energy-centric broadcast protocols in wireless sensor
  networks.
\newblock In Parosh~Aziz Abdulla and Carole Delporte{-}Gallet, editors, {\em
  Networked Systems - 4th International Conference, {NETYS} 2016, Marrakech,
  Morocco, May 18-20, 2016, Revised Selected Papers}, volume 9944 of {\em
  Lecture Notes in Computer Science}, pages 87--101. Springer, 2016.

\bibitem{LGS}
Kai Chen and Klara Nahrstedt.
\newblock Effective location-guided tree construction algorithms for small
  group multicast in {MANET}.
\newblock In {\em Proceedings of the 21st Annual Joint Conference of the {IEEE}
  Computer and Communications Society ({INFOCOM}-02)}, volume~3 of {\em
  Proceedings IEEE INFOCOM 2002}, pages 1180--1189, Piscataway, NJ, USA, June
  23--27 2002. IEEE Computer Society.

\bibitem{CFR}
Thomas Clouser, Mark Miyashita, and Mikhail Nesterenko.
\newblock Concurrent face traversal for efficient geometric routing.
\newblock {\em Journal of Parallel and Distributed Computing}, 72(5):627--636,
  2012.

\bibitem{ding2014algorithms}
Wei Ding and Ke~Qiu.
\newblock Algorithms for the minimum diameter terminal steiner tree problem.
\newblock {\em Journal of Combinatorial Optimization}, 28(4):837--853, 2014.

\bibitem{DHM14C}
Tony Ducrocq, Micha{\"{e}}l Hauspie, and Nathalie Mitton.
\newblock Geographic routing with partial position information.
\newblock In Octavian Postolache, Marten van Sinderen, Falah~H. Ali, and
  C{\'{e}}sar Benavente{-}Peces, editors, {\em {SENSORNETS} 2014 - Proceedings
  of the 3rd International Conference on Sensor Networks, Lisbon, Portugal, 7 -
  9 January, 2014}, pages 165--172. SciTePress, 2014.

\bibitem{DHMP14C}
Tony Ducrocq, Micha{\"{e}}l Hauspie, Nathalie Mitton, and Sara Pizzi.
\newblock On the impact of network topology on wireless sensor networks
  performances: Illustration with geographic routing.
\newblock In Leonard Barolli, Kin~Fun Li, Tomoya Enokido, Fatos Xhafa, and
  Makoto Takizawa, editors, {\em 28th International Conference on Advanced
  Information Networking and Applications Workshops, {AINA} 2014 Workshops,
  Victoria, BC, Canada, May 13-16, 2014}, pages 719--724. {IEEE} Computer
  Society, 2014.

\bibitem{Finn}
G.G. Finn.
\newblock Routing and addressing problems in large metropolitan-scale
  internetworks.
\newblock Technical Report ISI/RR-87-180, March 1987.

\bibitem{worldsensIPSN}
Antoine Fraboulet, Guillaume Chelius, and Eric Fleury.
\newblock Worldsens: development and prototyping tools for application specific
  wireless sensors networks.
\newblock In {\em Information Processing in Sensor Networks, 2007. IPSN 2007.
  6th International Symposium on}, pages 176--185. IEEE, 2007.

\bibitem{GS69}
K~Ruben Gabriel and Robert~R Sokal.
\newblock A new statistical approach to geographic variation analysis.
\newblock {\em Systematic Biology}, 18(3):259--278, 1969.

\bibitem{GJ79}
Michael~R Gary and David~S Johnson.
\newblock Computers and intractability: A guide to the theory of
  np-completeness, 1979.

\bibitem{steiner}
F.K. Hwang, D.S. Richards, and P.~Winter.
\newblock {\em The Steiner Tree Problem}, volume~53 of {\em Annals of Discrete
  Mathematics}.
\newblock North-Holland: Elsevier, 1992.

\bibitem{GPSR}
B.~Karp and H.T. Kung.
\newblock {GPSR}: Greedy perimeter stateless routing for wireless networks.
\newblock In {\em Proceedings of the Sixth Annual ACM/IEEE International
  Conference on Mobille Computing and Networking ({MobiCom})}, pages 243--254.
  ACM Press, August 2000.

\bibitem{TheoryPractice}
F.~Kuhn, R.~Wattenhofer, Y.~Zhang, and A.~Zollinger.
\newblock Geometric ad-hoc routing: Of theory and practice.
\newblock {\em 22nd ACM Symposium on the Principles of Distributed Computing
  (PODC)}, July 2003.

\bibitem{rectangleApproximation}
Marek Lassak.
\newblock Approximation of convex bodies by rectangles.
\newblock {\em Geometriae Dedicata}, 47(1):111--117, 1993.

\bibitem{NPHARD}
P.~C. Liu and R.~C. Geldmacher.
\newblock {On the deletion of nonplanar edges of a graph}.
\newblock In {\em Proc. of the 10th Southeastern Conference on Combinatorics,
  Graph Theory, and Computing}, pages 727--738, 1979.

\bibitem{PBM}
Martin Mauve, Holger F{\"u}{\ss}ler, J{\"o}rg Widmer, and Thomas Lang.
\newblock Position-based multicast routing for mobile ad-hoc networks.
\newblock {\em Mobile Computing and Communications Review}, 7(3):53--55, 2003.

\bibitem{improvedSteinerApprox}
Gabriel Robins and Alexander Zelikovsky.
\newblock Improved steiner tree approximation in graphs.
\newblock In {\em SODA}, pages 770--779. Citeseer, 2000.

\bibitem{T80}
Godfried~T Toussaint.
\newblock The relative neighbourhood graph of a finite planar set.
\newblock {\em Pattern recognition}, 12(4):261--268, 1980.

\bibitem{geosteiner}
David~M Warme, Pawel Winter, and Martin Zachariasen.
\newblock Exact solutions to large-scale plane steiner tree problems.
\newblock In {\em Proceedings of the tenth annual ACM-SIAM symposium on
  Discrete algorithms}, pages 979--980. Society for Industrial and Applied
  Mathematics, 1999.

\bibitem{GMP}
Shibo Wu and K.~Selcuk Candan.
\newblock {GMP}: Distributed geographic multicast routing in wireless sensor
  networks.
\newblock In {\em 26th IEEE International Conference on Distributed Computing
  Systems (26th ICDCS'06)}, page~49, Lisboa, Portugal, July 2006. IEEE Computer
  Society.

\end{thebibliography}
\end{document}